\documentclass[
 11pt, 
 preprint, 
 showkeys,
 superscriptaddress,
 nofootinbib,
 amsmath,amssymb,
 aps,
]{revtex4-2}

\usepackage{graphicx}
\usepackage{mathtools}
\usepackage{xfrac}
\usepackage{siunitx}
\usepackage{ulem}
\usepackage{amsmath}
\usepackage{multirow}
\usepackage{upgreek}
\usepackage{svg}
\usepackage{physics}
\usepackage{booktabs}
\usepackage[symbol]{footmisc}

\begin{document}
\pagenumbering{roman}

\title{Emergent functional dynamics of link-bots}

\author{Kyungmin Son}
 \thanks{These authors contributed equally to this paper}
 \affiliation{Department of Mechanical Engineering, Seoul National University, Seoul 08826, Korea}
\author{Kimberly Bowal}
 \thanks{These authors contributed equally to this paper}
 \affiliation{School of Engineering and Applied Sciences, Harvard University, Cambridge, MA 02138, USA}
\author{L. Mahadevan}
 \thanks{lmahadev@g.harvard.edu}
 \affiliation{School of Engineering and Applied Sciences, Harvard University, Cambridge, MA 02138, USA}
 \affiliation{Departments of Physics, and Organismic and Evolutionary Biology, Harvard University, Cambridge, MA 02138, USA}
\author{Ho-Young Kim}
 \thanks{hyk@snu.ac.kr}
 \affiliation{Department of Mechanical Engineering, Seoul National University, Seoul 08826, Korea}
 \affiliation{Institute of Advanced Machines and Design, Seoul National University, Seoul 08826, Korea}

\date{\today}

%--------------------------------------------------------------------
%     abstract
%--------------------------------------------------------------------
\begin{abstract} 
Synthetic active collectives, composed of many nonliving individuals capable of cooperative changes in group shape and dynamics, hold promise for practical applications and for the elucidation of guiding principles of natural collectives.
However, the design of collective robotic systems that operate effectively without intelligence or complex control at either the individual or group level is challenging. We investigate how simple steric interaction constraints between active individuals produce a versatile active system with promising functionality.
Here we introduce the link-bot: a V-shape-based, single-stranded chain composed of active bots whose dynamics are defined by its geometric link constraints, allowing it to possess scale- and processing-free programmable collective behaviors.
A variety of emergent properties arise from this dynamic system, including locomotion, navigation, transportation, and competitive or cooperative interactions. Through the control of a few link parameters, link-bots show rich usefulness by performing a variety of divergent tasks, including traversing or obstructing narrow spaces, passing by or enclosing objects, and propelling loads in both forward and backward directions. The reconfigurable nature of the link-bot suggests that our approach may significantly contribute to the development of programmable soft robotic systems with minimal information and materials at any scale.
\end{abstract}


\maketitle

%--------------------------------------------------------------------
%     summary figure and highlights
%--------------------------------------------------------------------
\newpage
\onecolumngrid

% empty line above necessary for centering
\begin{figure}[ht!]
    \label{OverviewFig}
    \centering
    \includegraphics{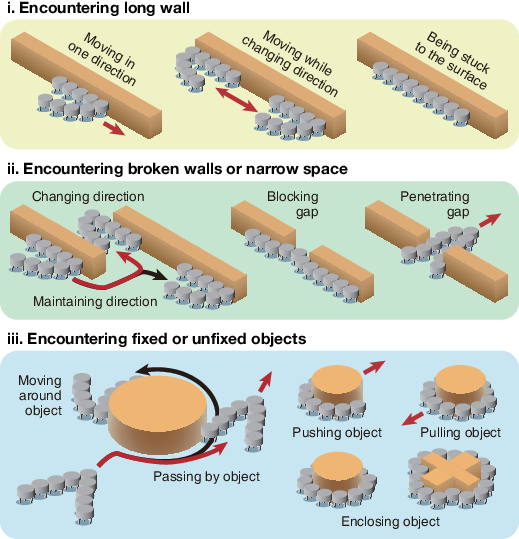}
    %\caption{\label{Overview} \textbf{Diverse behaviors and functions of link-bots.} Schematic illustrating the various functions performed by link-bots. The top panel demonstrates the link-bot's ability to move in one direction or move while changing direction along a wall, as well as adhere to the wall without motion. In the middle panel, the link-bot maintains or changes direction when encountering altered terrain, and either blocks or passes through narrow gaps. The bottom panel depicts the link-bot's capability to move around or passes through objects, and to push, pull, or completely encloses objects.}
\end{figure}
\vspace{1in}

\textbf{Highlights:}
\begin{itemize}
    \item The link-bot is a V-shaped chain of active bots connected by rigid links with rotational constraints
    \item Adjusting the link constraints provides simple and scale-free control of link-bot gaits
    \item In complex environments, link-bot gaits produce a rich variety of useful behaviors including locomotion, navigation, transportation, and interactions
    \end{itemize}

% %--------------------------------------------------------------------
% %     paper body
% %--------------------------------------------------------------------
% \twocolumngrid  % uncomment for reprint style
\clearpage
\cleardoublepage \pagenumbering{arabic} % start numbering pages from here

Active collectives composed of many individuals can cooperatively execute functions that are impossible for solo individuals to accomplish, e.g. complex architectures, predation escape or prey capture, brood care in social insects, etc. Synthetic systems with these properties provide an opportunity to address functional applications or elucidate guiding principles in natural collectives. Designing such systems is challenging, and efforts can be categorized into two approaches: an intelligent group is created by complex individuals programmed to work together ~\cite{rubenstein2014programmable, slavkov2018morphogenesis}, or group intelligence is an emergent property of simple individuals that spontaneously arises from their interactions ~\cite{giomi2013swarming}. The former approach is represented by macroscale swarm robotics, in which individuals equipped with sensing, memory, computation, and/or communication capabilities can perform useful group behaviors, such as constructing a target shape or migrating toward a specific destination~\cite{turgut2008self}. However, this approach has scale, computation, and communication limits because complexity is required at the individual level. These constraints are minimized in active particle systems that employ the second approach through the use of a stimulus such as light~\cite{lavergne2019group, vutukuri2020light, zhang2021cooperative}, acoustics~\cite{aghakhani2020acoustically}, or magnetic fields~\cite{yu2018ultra, wang2022order, ceron2023programmable} to promote desired collective behaviors such as locomotion, flocking, navigation~\cite{xie2019reconfigurable}, and transportation~\cite{gardi2022microrobot}. This is a useful approach, but it is limited by the requirement of an external global stimulus, which dictates the possible environment and materials as well as time and length scales.

A promising approach to circumvent these limitations involves generating collective behavior through physical interactions among active components, such as the flocking-like motion that emerges within granular shaken materials in the absence of any external control~\cite{narayan2007long}. Similar behaviors are seen in robot collectives through stochastic mechanical interactions that are adjusted at the group level through flexible and mobile boundaries~\cite{deblais2018boundaries, boudet2021collections, savoie2019robot} or through the type and strength of coupling between individuals~\cite{li2019particle, li2021programming, xi2024emergent}. The latter method of connecting individuals into a flexible chain or loop shows promise at the microscale~\cite{spellings2015shape,scholz2021surfactants,agrawal2020scale} and the macroscale~\cite{ozkan2021self,kulkarni2020reconfigurable}. 
The ability to alter the morphology of a collective system has also been shown to allow complex behaviors~\cite{friedl2009collective, vicsek2012collective} and adjustment to predetermined configurations~\cite{boley2019shape} at any scale. 

In this study, we investigate how simple steric interaction rules between active individuals produce a versatile active system with promising functionality by introducing the link-bot, a chain of forward-propelled bots defined by its internal geometric interaction constraints. In this active system, a few influential link parameters control the relative translation and rotation of each bot, allowing for breathing and flapping movements. These movements loosely control the link-bot shape and translate into predictable gaits when the link-bot encounters a boundary. When placed in complex environments, these morphological and translational movements produce a variety of emergent behaviors, including directed motion, interactions with obstacles, and transport of loads. The link-bot’s versatility is demonstrated by its ability to perform multiple contrasting functions: maintaining or changing direction in obstructive terrains, infiltrating or blocking narrow spaces, maneuvering past or around objects, and carrying objects forward or backward. Due to its scalability, material independence, and reconfigurability, this system paves the way for the development of functional, controllable, and autonomous collective systems using simple individuals at any scale.

\begin{figure}
\includegraphics[width=\textwidth, keepaspectratio]{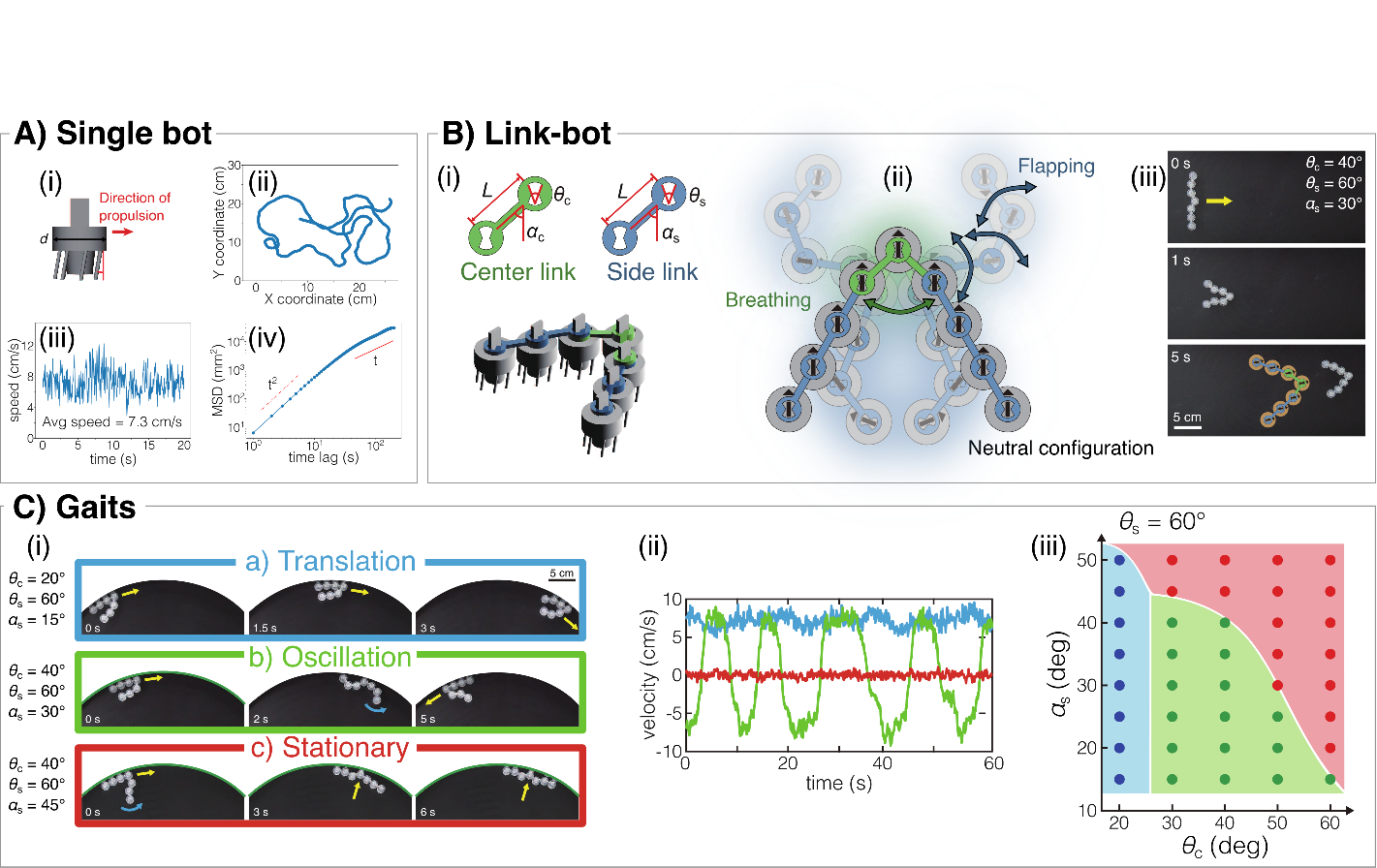}
\caption{\label{experiments} \textbf{The structure and dynamics of a single bot and link-bot in experiments.} \textbf{(A)} A single bot consists of (i) a cylindrical body on tilted legs and topped with a crest. The (ii) trajectory, (iii) speed profile, and (iv) diffusion plot of a single bot show characteristic active Brownian motion. \textbf{(B)} To construct a link-bot, rigid links connect $N$ bots together in a symmetric V-shape. The links have (i) length $L$, notch angle $\theta$, and spread angle $\alpha$, with two center links connecting the center bot at the V vertex and side links connecting all other bots that make up the side chains. In the neutral V-shape configuration (ii), all bot crests are aligned in the direction of motion. The link constraints allow two main modes of link-bot movement: breathing (shown by the green arrow) in which the central V angle opens and closes, and flapping (blue arrows) where the side chains bend inwards and outwards. (iii) Regardless of the initial configuration, the self-propelled link-bot relaxes into its neutral configuration (seen in the inset schematic) determined by the link properties. \textbf{(C)} Link-bots exhibit three gaits at a boundary, controlled by the link angles: translation (unidirectional motion, shown in blue), oscillation (changing directions by flipping along the wall, shown in green), and stationary (pushing against the wall without significant movement, shown in red). (ii) The changes in velocity for the three gaits. (iii) Phase diagram showing how the gaits change based on $\theta_\mathrm{c}$ and $\alpha_\mathrm{s}$.}
\end{figure}

\section{Link-bot structure}
Individual bots are 3D printed, consisting of a cylindrical body (diameter, $d = 1.5$ cm) on seven circumferentially equidistant legs, pictured in Fig.~\ref{experiments}A(i). The legs are tilted, allowing the bot to self-propel in a preferred direction when placed on a vibrating surface. A circular flat arena of diameter 45 cm is vertically vibrated at a frequency of $\approx 80$ Hz and an amplitude of 70 $\upmu$m, causing a single bot to move at an average speed of 8 $\mathrm{cm}/\mathrm{s}$. The arena vibration properties are kept constant in all experiments. Fig.~\ref{experiments}A shows an example trajectory (ii) and speed profile (iii) of a single bot moving freely for 20~s. The corresponding plot of the translational mean squared displacement with respect to time lag is given in Fig.~\ref{experiments}A(iv), which shows ballistic motion ($\sim t^2$) over short time scales and diffusive motion ($ \sim t$) over large time scales, typical of active Brownian motion.

Bringing multiple bristle-bots into a collective provides rich and interesting behavior~\cite{giomi2013swarming, hao2022controlling}. Previous work on bristle-bots that are connected to form an active chain focuses on elasto-active systems~\cite{xi2024emergent, xu2024constrained, zheng2023self} and the mechanical coupling of connected active chains~\cite{xia2024biomimetic, zheng2023self}. In this work, we focus on systems of bots that are connected by rigid links with rotational constraints: link-bots. The link-bot is created by connecting $N$ bots with $N - 1$ links in a V-shaped arrangement, inspired by the formations observed in troops and migrating birds~\cite{bajec2009organized}. An example where $N=7$ is pictured in Fig.~\ref{experiments}B. Each bot has a cuboidal crest on its top surface, which allows it to fit into the ribbon-shaped notches on both ends of the links. These links serve to maintain constant interbot distances between neighbors, transmit the motion of each bot to its neighbors and constrain each bot's rotation. Links are characterized by three parameters: length between notches $L$, notch angle $\theta$, and spread angle $\alpha$. The center bot and its neighbors are connected using two center links (pictured in green in Fig.~\ref{experiments}B(i)), while all other bots are connected with side links (pictured in blue). The center and side links always have the same length $L=1.6$ cm, although their angles may differ and will be reported using subscripts c and s for the center and side, respectively. To produce a V-shaped arrangement, the links on one side of the center bot are reversed in relation to the links on the other side. This symmetry effectively suppresses undesired random deformations, such as crumpling and curling, which are often observed in active filaments~\cite{winkler2017active}. Two notable features of the link-bot in comparison to previous connected bristle-bot systems are the broken asymmetry introduced by the V-shape and the threshold constraints imposed by the link notch angles. These features allow for a rich variety of collective behaviors to emerge from its characteristic active chain dynamics to provide a multi-functional soft robotic system.

\section{Model}
\begin{figure}
\includegraphics[width=\textwidth, keepaspectratio]{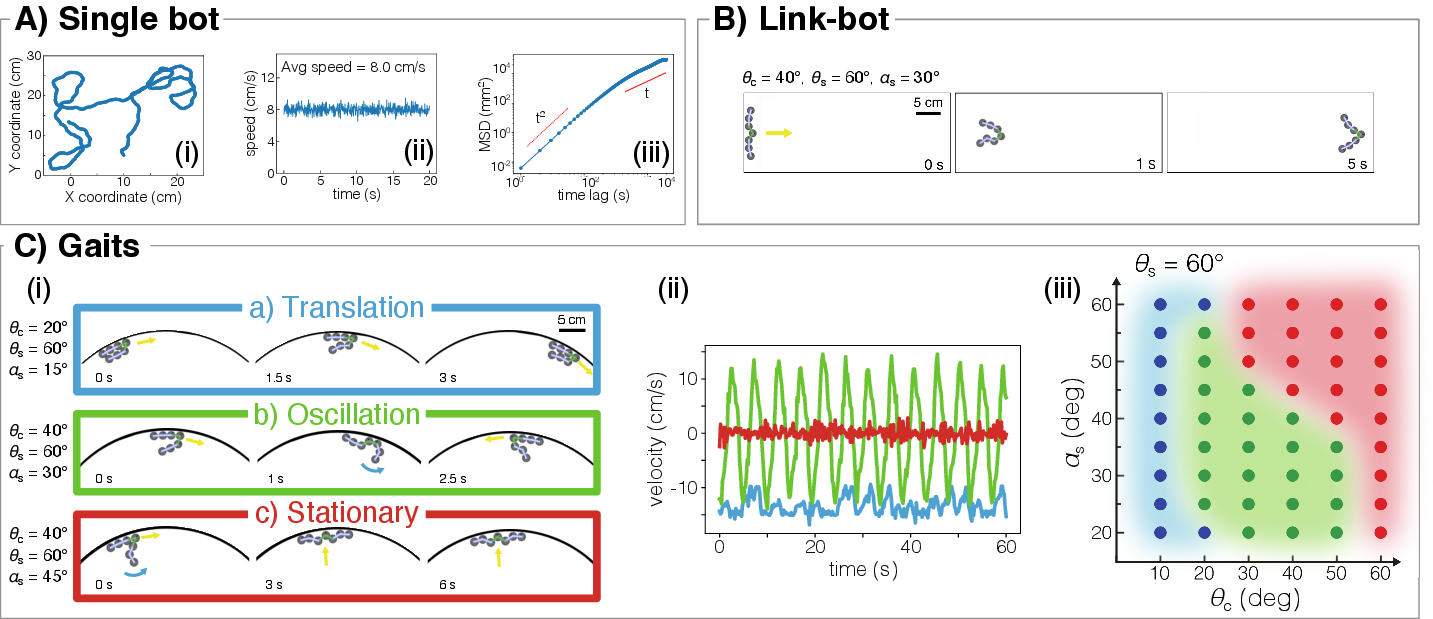}
\caption{\label{simulations} \textbf{Link-bot structure and dynamics in the computational model.} \textbf{(A)} The (i) trajectory, (ii) speed, and (iii) MSD of a single bot over 20~s is pictured, showing active Brownian motion matching that of the experiments. \textbf{(B)} Each self-propelled bot within a link-bot is modeled with translational and rotational constraints caused by the center and side links. This results in a noisy relaxation to the neutral configuration, where the bots form a V with all crests aligned, when the link-bot moves forward in space, regardless of the initial configuration. \textbf{(C)} (i) As in the experiments, the modeled link-bots exhibit three gaits at a wall: translation, oscillation, and stationary. (ii) The gaits are distinctive in their velocity patterns and (iii) show the same dependencies on $\theta_\mathrm{c}$ and $\alpha_\mathrm{s}$ at $\theta_\mathrm{s}=60^\circ$ as seen in experiments.}
\end{figure}
A computational model is developed to thoroughly evaluate the dynamical behaviors of the link-bot, explore the parameter space, and make predictions. Each bristle-bot is modeled as an active Brownian particle, moving due to self-propulsion and diffusion. An example 20~s trajectory for a single simulated bot is shown in Fig.~\ref{simulations}A(i), with a corresponding speed profile (ii) and diffusion plot (iii). The link-bot is simulated by adding translational and rotational constraints caused by the side and center links connecting the bots. Further details about the model are provided in Section~\ref{sec:model_eqns}.

\section{Locomotion} %link-bot dynamics
When not acted upon by outside forces, such as walls or obstacles, the link-bot moves forward in the direction of the center bot (i.e.\ the V vertex), generally maintaining a neutral configuration where all bots point in the same direction. The dimensions of this neutral configuration, examples of which are given in Fig.~\ref{experiments}B and Fig.~\ref{simulations}B, are controlled by the link lengths and angles, as well as the number of bots. 

The link-bot possesses two dynamic configuration modes due to the constraints imposed by the two link types: breathing and flapping. Breathing, shown by green arrows in Fig.~\ref{experiments}B(ii), occurs when the central angle opens and closes between its minimum, which is controlled by the steric interactions of the bots, and its maximum value of $\theta_\mathrm{c} + 2\alpha_\mathrm{c}$. The angle $\alpha_\mathrm{c}$ is not independent and is determined by the dimensions of the link and the bot as $\alpha_\mathrm{c} = \sin^{-1}[d/(2L)] + \theta_\mathrm{c}/2$. This means that the range of breathing movements of the link-bot is controlled only by $\theta_\mathrm{c}$.
Flapping, shown by blue arrows in Fig.~\ref{experiments}B(ii), occurs when the side chains bend outward or inward. This flagella-like movement is known to be exhibited by active particle chains when they are pinned at one end~\cite{chelakkot2014flagellar, zheng2023self, xu2024constrained}. The asymmetry introduced by the V-shape of the link-bot allows this behavior to emerge without external pinning. The rotational notch angle $\theta_\mathrm{s}$ and the side chain spread angle $\alpha_\mathrm{s}$ contribute in similar ways to the freedom of movement of a side bot. High values of $\theta_\mathrm{s}$ and $\alpha_\mathrm{s}$ allow for large flapping modes, producing floppy link-bot side chains that are able to bend significantly. In contrast, low values of these angles reduce flapping movement and produce rigid side chains that do not easily deform from the neutral configuration. These structural changes also translate into similar link-bot movement behaviors, which means that $\theta_\mathrm{s}$ and $\alpha_\mathrm{s}$ have the same phenotypic effects on the link-bot, as shown in Fig.~S6. Therefore, for simplicity unless otherwise stated, in the following work $\theta_\mathrm{s}$ is kept at 60$^\circ$, which provides a balance between the angle constraint and the freedom of bot motion. Therefore the side link spread angle $\alpha_\mathrm{s}$ controls the flexibility of the side chains and the resulting flapping movements.

\section{Gaits}
To see how the link-bot internal geometric parameters yield diverse behaviors, we investigate the link-bot's response upon encountering a wall. When the links do not impose any angular constraints on the bots, i.e.\ the link angles are set at $180^\circ$, the link-bot exhibits no directed motion or consistent interactions with the wall (shown in Fig.~S2 and Video~S1). Simulations show that smaller link angles enhance the coordinated collective behavior of the link-bot. Fig.~\ref{simulations}C shows the behaviors of a link-bot consisting of $N=7$ bots connected with different link angle values. The behavior can be categorized into three gaits: (i) unidirectional translation, (ii) oscillatory motion in which the link-bot periodically changes direction along the wall, and (iii) stationary. 
Fig.~\ref{simulations}C(ii) shows the velocity of the center bot as a function of time in each gait. The link-bot maintains a constant velocity when in translation, is periodic with a constant amplitude and frequency in the oscillation gait, and fluctuates around zero when stationary. The detailed dynamics of these gaits and how they can be understood by the torque applied around the center bot are discussed further in Fig.~S3 and Supplementary Information Section~3.
A phase diagram showing the link-bot gaits at a wall as a function of $\theta_\mathrm{c}$ and $\alpha_\mathrm{s}$ is shown in Fig.~\ref{simulations}C(iii). In line with the breathing and flapping modes, we see that the gait phenotypes are largely predicted by a small subset of the link-bot geometric parameters: the central angle $\theta_\mathrm{c}$ and the side chain flexibility $\alpha_\mathrm{s}$ (detailed discussion found in the Supplementary Information).  The experimental observations of these gaits, shown in Fig.~\ref{experiments}(C) and Video~S1, agree well with those predicted by the model. The link-bot length, controlled by $L$ and $N$, are seen to have a weak effect on gait (Figs.~S4 and S5, Video~S2), which means that flexibility and $\theta_\mathrm{c}$ are sufficient to predict gait (Fig.~S7).

\section{Navigation} %Multiple walls
\begin{figure}
\includegraphics[width=0.7\textwidth, keepaspectratio]{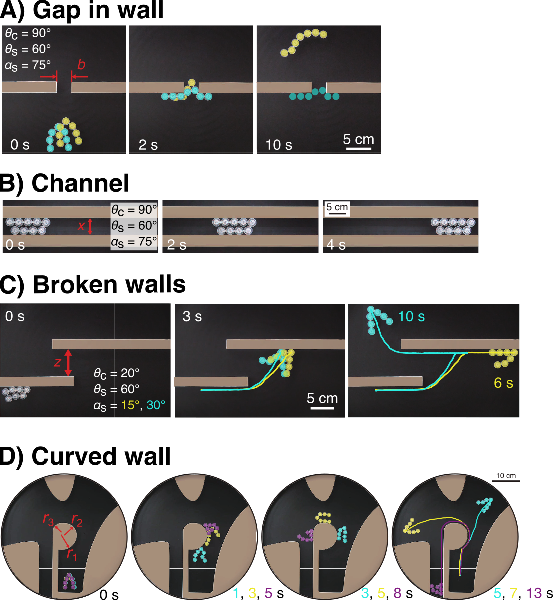}
\caption{\label{atwalls} \textbf{Link-bots possess programmable locomotion in a variety of environments.} \textbf{(A)} The trajectories of two identical link-bots through a wall with a gap of $b=3d$, showing that the alignment (yellow) or misalignment (blue) of the center bot with the gap plays an important role in determining whether the link-bot will pass through or remain stuck at the wall. \textbf{(B)} A link-bot traveling along a narrow channel with width $x=2d$. The speed and direction of the link-bot motion is controlled by the link angles. \textbf{(C)} Upon encountering two walls with a spacing of $z=4d$, link-bots with different side link angles either pass through the maze of walls (blue) or remain trapped at a wall (yellow). \textbf{(D)} The link angles of a link-bot control the distance it will travel along a curved surface. An obstacle with three curvatures ($r_1/d = 3.5, r_2/d = 2.8, r_3/d = 2$) can thus be used to sort link-bots of different properties. Shown here are link-bots with $\theta_\mathrm{c} = 20^\circ$, $\alpha_\mathrm{s} = 15^\circ$ (blue), $\theta_\mathrm{c} = 40^\circ$, $\alpha_\mathrm{s} = 30^\circ$ (yellow), and $\theta_\mathrm{c} = 20^\circ$, $\alpha_\mathrm{s} = 45^\circ$ with the left-most side link at 0~s inverted (purple).}
\end{figure}
When put in complex environments, link-bots in experiments are able to navigate in distinct exploratory or exploitative ways, as seen in Fig.~\ref{atwalls}. These contradictory behaviors are products of the link-bot gait and its effect on wall interactions in each case, and thus can be controlled by the link-bot angles. 
Exploratory behaviors, characterized by the link-bot traversing throughout its surroundings, occur when the link-bot not significantly constrained by its interactions with a boundary. Examples of exploratory behavior shown in Fig.~\ref{atwalls} are passing through a gap in a wall (A), traveling quickly through a channel (B), going around broken walls (C), and leaving a curved surface (D). Exploratory movement is generally favored by small to intermediate $\theta_\mathrm{c}$ and $\alpha_\mathrm{s}$ values which produce an oscillatory gait and keep bot self-propulsion tangential to gaps, channel walls, and curved surfaces.
In the example with discontinuous parallel walls (Fig.~\ref{atwalls}C), the exploratory link-bot, shaded in blue, navigates around the edges to continue past wall segments. This behavior is controlled by adjusting $\alpha_\mathrm{s}$ to produce intermediate side chain flexibility or changing $\theta_\mathrm{c}$ to generate an oscillatory gait (for more details regarding the probabilities of these dynamics in the noisy system, see Fig.~S12). In an environment with curved boundaries, decreasing the link-bot flapping and breathing modes, through the use of lower $\alpha_\mathrm{s}$ and $\theta_\mathrm{c}$ values, causes the link-bot leave the surface and explore its surroundings.

In contrast, localized exploitative behaviors, such as blocking a gap in a wall (Fig.~\ref{atwalls}A), traveling slowly through a channel (B), getting stuck at broken walls (C), or following a curved surface (D), dominate when the link-bot remains relatively stationary. Depending on the environment, this occurs at large or small values of the link angles. When rotational constraints are weak (i.e.\ $\theta_\mathrm{c}$ or $\alpha_\mathrm{s}$ are large) the resulting large breathing and flapping movements reach a stable state along straight surfaces where the bots are pushing against the walls so that normal or frictional forces restrict movement. 
In an environment with multiple walls, setting the link constraint angles to small values will produce an exploitative link-bot that does not pass through a maze of walls since it moves in the translation gait and is unable to change directions when encountering a new boundary (see the yellow link-bot in Fig.~\ref{atwalls}C). A link-bot with large link angles will possess a stationary gait that causes it to remain fixed at a wall and therefore fail to progress through a maze. Similarly, when a link-bot enters a narrow channel, its movement is strongly constrained by the parallel surfaces. A high $\alpha_\mathrm{s}$ value allows free rotation of bots towards both walls, increasing the active forces pointing into the walls and causing the link-bot speed to decrease (geometric prediction compared to experiments in Fig.~S10, Video~S4). This behavior is independent of link-bot length for $N = 7, 9, 11$ (Fig.~S10C). 

The position of the link-bot when encountering small boundary features can be an important factor for subsequent behaviors. For example, when the approaching center bot does not align with a gap in the wall, the link-bot sometimes will not bend inward and the resulting high resistance from the narrow gap and the adjacent wall prevents its passage through the gap, causing highly exploitative behavior (blue shaded link-bot in Fig.~\ref{atwalls}A, also shown in Video~S3).
To allow further control, an asymmetry can be added to the link-bot, such as the inversion of one of the end side links. This simple adjustment induces one-sided inward propulsion (Fig.~S13E), thereby enabling the link-bot to rotate around obstacles with significant curvature without moving away from the surface (Fig.~S13F and Video~S4). This expands the range of link-bot behaviors in this environment, allowing for an effective self-sorting mechanism such as that shown in Fig.~\ref{atwalls}D for three link-bots around one wall of varying curvatures.

These few barrier environments shown in Fig.~\ref{atwalls} can be extended to an arbitrary number of walls for increasingly complex and realistic environments, such as a building layout or maze, through which link-bot behavior can be controlled using its internal geometric constraints. More details and examples are shown in Figs.~S8, S9, S11, and Videos~S3--5.

\begin{figure}
\includegraphics[width=\textwidth, keepaspectratio]{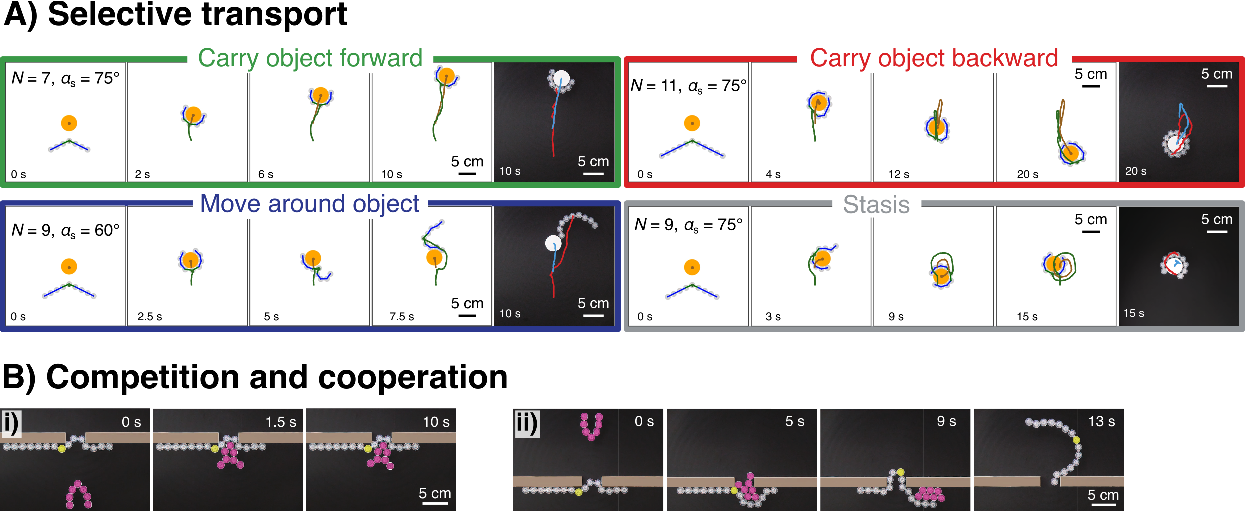}
\caption{\label{objects} \textbf{Link-bots perform selective transportation and dynamic social interactions}. \textbf{(A)} By adjusting the link angles and number of bots within a link-bot, its transport behaviors can be controlled. Snapshots show the following behaviors of the simulated link-bot when it encounters a passive mobile object: pushing the object forwards, pulling the object backwards, passing without carrying the object, and wrapping/rotating around the object staying relatively stationary. The link-bot trajectory is shown in green, the object trajectory is brown. The last snapshot in each panel with a black background shows the analogous experimental results at the final time point, with the link-bot trajectory shown in red and the object trajectory in blue. In all cases the diameter of the object is $d_{\mathrm{obj}}=2.67d$. \textbf{(B)} Two interacting experimental link-bots, one with $N=15$, $\theta_\mathrm{c}=90^\circ$, $\alpha_\mathrm{s}=75^\circ$ (colored gray, with the center bot shown in yellow) and another with $N=7$, $\theta_\mathrm{c}=20^\circ$, $\alpha_\mathrm{s}=15^\circ$ (colored pink), show competitive and cooperative actions in traversing a wall with a gap. In both cases, the longer link-bot initially blocks a gap of $2.67d$ in the wall. (i) In the competitive situation, a link-bot approaching from the same side of the wall encounters the blocking link-bot and gets stuck so that neither link-bot is able to move. (ii) In the cooperative case, the blocking link-bot is moved by the smaller link-bot which approaches from the other side of the wall. This motion allows both link-bots to traverse the gap and pass through the wall in opposite directions.}
\end{figure}
\section{Transportation and interactions}
The link-bot model is a useful tool to further explore link-bot functionality in complex scenarios in order to predict and design link-bot behaviors. We present simulation results showing how link-bots balance directed motion and structural flexibility to interact with mobile objects in useful ways.
Adjusting the geometric properties of the link-bot produces many different transportation behaviors, some of which are shown in Fig.~\ref{objects}. Here the center link angle is kept at $\theta_\mathrm{c} = 90^\circ$, so the breathing movement is minimally constrained. This allows the behavior to be controlled by $\alpha_\mathrm{s}$ and $N$ only, although it should be noted that low values of $\theta_\mathrm{c}$ will reduce the allowed breathing mode angle and cause the same effect as low $\alpha_\mathrm{s}$ values by reducing the link-bot contact with the object. 
When the link-bot is relatively short and the object is large, the link-bot carries the object forward. As the link-bot length increases, it is more likely to move around and away from the object. Link-bots interacting with relatively small objects are likely to carry the object backwards. At some intermediate values, the link-bot remains wrapped around the object in a state of stasis, with no significant translational movement. These trends are similar across angular constraints, with carrying behaviors enhanced by high side chain flexibility at high $\alpha_\mathrm{s}$ values and object avoidance behaviors promoted by rigid side chains at low $\alpha_\mathrm{s}$ values (phase diagrams and momentum analysis shown in Fig.~S16). These selective transportation behaviors predicted by the simulated link-bot are observed in experimental studies, shown as complementary final time point snapshots in Fig.~\ref{objects}A, and in more detail in Fig.~S15 and Video~S4.  The link-bot is not limited to interactions with a circular object. With sufficiently large $N$, $\theta_\mathrm{c}$, and $\alpha_\mathrm{s}$, a link-bot can enclose objects of diverse shapes, including ellipses, squares, triangles, L-shapes and cross shapes (Fig.~S17 and Video~S4).

To explore competitive and cooperative behaviors experimentally, we consider how two link-bots hinder or promote movement through a gap in a wall. In this scenario a long link-bot, colored gray in Fig.~\ref{objects}B, is engaged in an exploitative stationary gait at the gap when a second link-bot (colored pink) approaches from either the opposite or the same side of the wall. In the competitive case (i), both link-bots push into the wall together and both become jammed at the gap. In the cooperative case (ii), the pink link-bot is able to overcome the self-propulsion forces of the gray one and helps both pass through by realigning the gray link-bot with the gap. More details are shown in Fig.~S9.

\section{Discussion}
In summary, we present a collective system of simple bots that can locomote, navigate, transport, and interact in a variety of environments. By manipulating a few internal geometric constraints, these link-bots show functionality and versatility that have been a challenge to achieve in traditional robotic swarms without sophisticated control. This is explained by the ways the deformable active structure of the link-bot allows for breathing and flapping movement modes, which produce three gaits at a boundary. Advanced behavior does not require increased complexity at the link-bot level, but comes from complexity in the environment. 
We emphasize that the link-bot is able to perform a wide variety of functions with minimal control, with even more behaviors to be explored. This is shown through a range of contrasting tasks including navigating through or circumventing obstacles, adhering to or detaching from objects, transporting objects in forward or backward directions, traversing or blocking small gaps, allowing or obstructing the passage of objects through gaps, and self-sorting on a curved surface (Fig.~S18 and Video~S5 show additional examples where these behaviors are exhibited in a multicomponent environments).

Further work could extend the capabilities of the link-bot by making the links and crests dynamically adjustable. The ability to modify link and crest shapes and material properties, using environmental conditions for example, would allow for a broader range of movements and remotely controlled on-the-fly transformations. Leveraging its potential for extensibility, the link-bot principle of using geometric constraints between active bots can serve as a starting point for developing versatile and minimalistic robot collectives across various scales. These principles could also be applied to develop robust and cost-effective robots for tasks such as transporting goods over challenging terrain, conducting environmental surveillance, or controlling traffic flow. Our findings provide valuable insight into the development of multifunctional robotic systems that are both resource-efficient and scalable, with the potential to impact a wide range of industries and activities.
\clearpage

\section{Methods}
\subsection{Experiments}
\noindent\textbf{Link-bot fabrication:} As shown in Fig.~\ref{experiments}A, each bot comprises two cylinders, the cap (diameter $d$ 15~mm, height 6.5~mm) and the body (diameter 8~mm, height 6.5~mm), and a cuboid top (width 1.5~mm, length 5~mm, height 8~mm) on the cap, all three of which are connected on the same axis. The cap is equipped with seven legs (length 8 mm, diameter 1 mm) that are tilted at an angle of 10$^\circ$ from the vertical direction. The connecting link (thickness 1.5~mm) consists of two disks (diameter 8.5~mm) joined by a bar (length $L =$ 16~mm, width 2.5~mm). The length $L$ slightly exceeds $d$, allowing the creation of a chain without direct contacts between adjacent bots. Each disk features a ribbon-shaped notch that allows the cuboid crest of the bot to rotate freely within a prescribed angle, and the two notches in the link have the same orientation. To fabricate all bots and links, a transparent photopolymer with an acrylate base is used, employing stereolithography 3D printing which has an accuracy of approximately $\pm 0.1$~mm (Formlabs Form 3). For the analysis of the link-bot's travel distance, velocity, and trajectory, we trace and examine the position of the center bot using the TrackMate plug-in for ImageJ~\cite{tinevez2017trackmate}.

\noindent\textbf{Vibrating table:} The bots are subjected to excitation through the vertical vibration of a circular acrylic base plate (diameter 480~mm, height 30~mm), firmly mounted on an electromagnetic shaker (Tira TV 5220). The plate is maintained in a horizontal position with a precision of 0.1$^\circ$. The motion of the bots is confined within a circular boundary (diameter 450~mm). To mitigate resonance effects, the shaker is attached to massive concrete blocks. Experiments are carried out with vibrations at a frequency of $80~\mathrm{Hz}$ and an amplitude of $70~\upmu$m, ensuring a consistent and steady excitation of the bots.

\subsection{Computational model}
\label{sec:model_eqns}
\begin{figure}
\includegraphics[width=0.9\textwidth, keepaspectratio]
{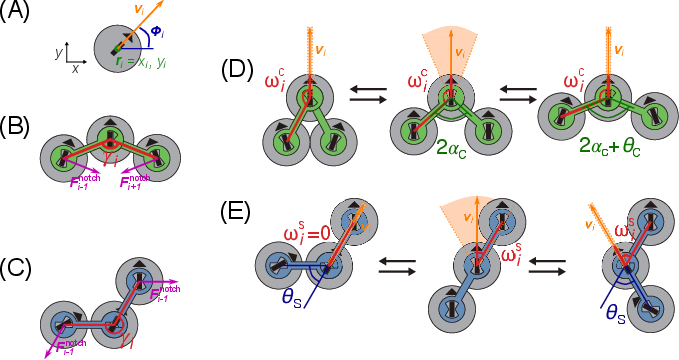}
\caption{\label{Methodsfig} \textbf{Schematic representation of the angles and vectors in the link-bot model}. \textbf{(A)} Position and velocity vector of a single bot. Schematics of a partial link-bot showing the translational forces, $\mathbf{F}^{\mathrm{notch}}$, due to the links for \textbf{(B)} bots connected by center links and \textbf{(C)} bots connected by side links. Pictured here is a side chain to the left of the center bot, which is flapping outward. For the case where the side chain flaps inward, the results are mirrored. For both the center and side links, $\mathbf{F}^{\mathrm{notch}} > 0$ only when $\gamma_{i} > \gamma_{\mathrm{max}}$. The notches constrain bot rotation for \textbf{(D)} the center bot and \textbf{(E)} the side bots. In the neutral configuration (shown in the middle) the bots possess maximum rotational freedom, shown by a shaded orange region. When the bots are in their fully extended breathing and flapping modes (shown for both directions on either side) the bots have no rotational freedom.}
\end{figure}

\noindent\textbf{Bot activity:} A model was created using Python to simulate link-bot behavior in order to investigate detailed properties and extend the parameter scope. The link-bot is modeled as a collective of $N$ active Brownian particles interacting through constraints imposed by the connecting links and the surrounding environment. 
Each circular bot $i$ has position $r_{i}$ and orientation $\phi_{i}$ at time $t$, shown in Fig.~\ref{Methodsfig}A, which are updated according to the following dynamical update rule:
\begin{equation}
    \begin{split}
    \mathbf{v}_{i}^{t} = \mathbf{v}_{i}^{t-1} &+ \Bigg[ \sum_{i=1}^{N-1} \underbrace{\mathbf{F}_{i,j=i+1}^{\mathrm{overlap}}}_\text{forces due to bot-bot interactions} \\
    &+ \sum_{i=2}^{N-1} \underbrace{(\mathbf{F}_{i,j=i+1}^{\mathrm{link}} + \mathbf{F}_{i,j=i-1}^{\mathrm{link}}) + \mathbf{F}_{i=1,j=2}^{\mathrm{link}} + \mathbf{F}_{i=N,j=N-1}^{\mathrm{link}}}_\text{forces due to rigid link length} \\
    &+ \sum_{i=2}^{N-1} \underbrace{(\mathbf{F}_{i, j=i+1}^{\mathrm{notch}} + \mathbf{F}_{i,j=i-1}^{\mathrm{notch}}) + \mathbf{F}_{i=1,j=2}^{\mathrm{notch}} + \mathbf{F}_{i=N,j=N-1}^{\mathrm{notch}}}_\text{forces due to link notch constraints} \Bigg] \cdot \mathrm{d}t 
    \end{split}
\end{equation}
where the velocity of the bot at time $(t-1)$ is given by

\begin{equation}
    \label{particle_v_eqn}
    \mathbf{v}_{i}^{t-1} = v_{0} \left[ \begin{array}{c} \cos \phi_{i} \\ \sin \phi_{i} \end{array} \right] + \sqrt{2D} \cdot \eta_{i}
\end{equation}
Here, bot activity is given by a constant self-propulsion speed $v_{0}$, the diffusion coefficient $D$, and a noise array randomly sampled from standard normal distribution $\eta_{i}$ (Equation~\ref{particle_v_eqn}). More information about how $D$ is calibrated from experiments is given in Fig.~S1. We observe experimentally that the bots move primarily along via translation, and do not rotate without translation, i.e. they are subject to a nonholonomic constraint~\cite{Bloch-Nonholonomic-mechanics}. This coupling of the orientational movement to translation is likely due to the fact that the bot motion is generated tangential to the leg tilt direction on the vibrating surface. Therefore, the noise in the model is applied to translation of the bot rather than rotation, which is seen to provide appropriate dynamics. Some quantitative differences can be observed in freely moving link-bots if noise is applied to both rotation and translation (for example, in the relaxation to the V-shaped neutral configuration), but no significant differences were seen in the gaits and dynamics reported in this work.

\noindent\textbf{Link constraints:} The forces acting on each bot due to its connecting link(s) and any external constraints are implemented using the geometry of the link-bot with respect to the position of the bots. The rigidity of each bot and the connecting links are maintained through linear spring constraints. %elastic forces
\begin{equation}
    \mathbf{F}_{i,j}^{(\cdot)} = -k(\mathbf{r}_{i,j} - r_{0}) \cdot \hat{\mathbf{r}_{ij}},
\end{equation}
where $\mathbf{F}_{i,j}^{\mathrm{\mathrm{overlap}}}$ uses $r_{0} = L$ and $\mathbf{F}_{i,j}^{\mathrm{link}}$ uses $r_{0} = d$.

All spring constants are set to a sufficiently high value, $k=\num{2e5}$. Wall and object boundary constraints are implemented as perfectly elastic collisions.

The link notches are a key feature of the link-bot, providing a hard boundary that affects both bot translation and rotation. These constraints cause each bot's movement to be coupled with the relative positions of its neighboring bots. This is enhanced by the fact that all bots, except for the side bots at the end of the chains, are controlled by two overlapping links. This feature contributes to the emergent complex behaviors of the link-bot. 

The translational constraints imposed on each bot by the link notches are implemented as an exponential spring:
\begin{equation}
    \mathbf{F}_{i,j}^{\mathrm{notch}} = -k \cdot \exp(\gamma_{i} - \gamma_{\mathrm{max}}) \cdot \mathcal H (\gamma_{i} - \gamma_{\mathrm{max}}),
\end{equation}
where $\gamma_{i}$ is the angle made by the two links connected to bot $i$ and $\gamma_{\mathrm{max}}$ is a maximum angle constraint set by the notches. For the center bot, $\gamma_{\mathrm{max} } = 2\alpha_\mathrm{c} + \theta_\mathrm{c}$, and for the side bots, $\gamma_{\mathrm{max}} = 180^\circ + \theta_\mathrm{s}$. The Heaviside step function $\mathcal H(\gamma_{i} - \gamma_{\mathrm{max}})$ ensures that the bots move freely except when the limits of the notch are reached. Fig.~\ref{Methodsfig}B and C provide schematics of these angles and forces for sample partial link-bots.

The rotational boundaries felt by each bot due to its one or two overlapping link notches are implemented as hard angle constraints on the bot velocity vector. Equations~(\ref{eqn:omega_centre}- \ref{eqn:omega_side}) show these clamped angle limits for center and side bots when situated in a neutral configuration with their neighbors (i.e.\ all crests are aligned), as shown in the middle schematic of Fig.~\ref{Methodsfig}D and E.
\begin{equation}
    \label{eqn:omega_centre}
    \omega_{i}^\mathrm{c} = \max \{ 180^\circ - \alpha_\mathrm{c} - \theta_\mathrm{c}/2, \min(\omega_{i}^\mathrm{c}, 180^\circ - \alpha_\mathrm{c} + \theta_\mathrm{c}/2) \} ,
\end{equation}
\begin{equation}
    \label{eqn:omega_side}
    \omega_{i}^\mathrm{s} = \max \{ \alpha_\mathrm{s} - \theta_\mathrm{s}/2, \min(\omega_{i}^\mathrm{s}, \alpha_\mathrm{s} + \theta_\mathrm{s}/2) \} ,
\end{equation}
where $\omega_{i}^\mathrm{c}$ is the angle between the center bot crest and its neighboring link to the left and $\omega_{i}^\mathrm{s}$ is the angle between the side bot crest and its neighboring link above (closer to the center bot). When the link-bot engages in breathing or flapping movements, the notches on the bots move in opposite directions and lead to increased constraint of the bot rotation. This is shown in Fig.~\ref{Methodsfig}D and E in the minimally and maximally extended cases in which the bot has no rotational freedom. For example, when the link-bot center V-angle is at its maximum value of $2\alpha_\mathrm{c} + \theta_\mathrm{c}$, the center bot's orientation is completely constrained since the two notches surrounding its crest are fully rotated in opposite directions. In this situation the center bot cannot rotate and points straight forward relative to the link-bot V-shape ($\omega^\mathrm{c} = 180^\circ - \alpha_\mathrm{c} - \theta_\mathrm{c}/2$).

The main parameters of the model and their typical values are provided in Table~\ref{parameter_table}.
\begin{table}[h]
\caption{\label{parameter_table} \textbf{Parameters used in link-bot model with typical values}}  
\begin{tabular}{cccc} \toprule
    \textbf{Parameter} & \textbf{Description} & \textbf{Typical Value(s)} & \textbf{Unit} \\ \midrule
    $d$ & bot diameter & 1.5 & cm \\
    $v_{0}$ & bot self-propulsion speed & 8 & cm/s \\ 
    $D$ & bot diffusion coefficient & \num{1e3} & $\mathrm{cm}^2/\mathrm{s}$ \\ %\midrule
    $L$ & link length & 1.6 & cm \\
    $\theta_{\mathrm{c}}$ & center link notch angle & 10 -- 180 & deg \\
    $\alpha_{\mathrm{c}}$ & center link spread angle & 10 -- 90 & deg \\
    $\theta_{\mathrm{s}}$ & side link notch angle & 10 -- 180 & deg \\
    $\alpha_{\mathrm{s}}$ & side link spread angle & 10 -- 90 & deg \\ 
    $k$ & spring constant & \num{2e5} & N/m \\
    $N$ & number of bots in a link-bot & 3 -- 31 & -- \\
    $b$ & gap in wall spacing & 3 -- 6 & cm \\
    $x$ & channel spacing & 3 & cm \\
    $z$ & broken wall spacing & 3 -- 7.5 & cm \\
    $r$ & radius of curved wall & 3 -- 15 & cm \\
    $d_{\mathrm{obj}}$ & object diameter & 1.5 -- 15 & cm \\  \bottomrule
\end{tabular}
\end{table}
\clearpage

\begin{acknowledgments}
\textbf{Funding:} This work was supported by the National Research Foundation of Korea (Grant Nos.~2018-052541 and 2021-017476) via the SNU SOFT Foundry Institute. H.-Y.K. acknowledges administrative support from SNU Institute of Engineering Research. K.B. acknowledges support from the Human Frontier Science Program grant LT000444/2021-C.
\textbf{Author contributions:} Conceptualization: K.S., K.B., H.-Y.K., L.M. Experiments: K.S. Computations: K.B.  Analysis and Interpretation: K.B., L.M., K.S., H.-Y.K. Visualization: K.B., K.S. Supervision: H.-Y.K., L.M. Writing--original draft: K.S., K.B. Writing--review and editing: K.B., H.-Y.K, L.M.
\textbf{Competing interests:} The authors declare that they have no competing interests.
\textbf{Data and materials availability:} All data are available in the manuscript and the supplementary materials.
\end{acknowledgments}

% %--------------------------------------------------------------------
% %     bibliography
% %--------------------------------------------------------------------
% \clearpage \citeindexfalse

\bibliography{bibliography}    %Create bibliography

% %--------------------------------------------------------------------
% %     appendices
% %--------------------------------------------------------------------
\clearpage

\end{document}